\documentclass{czjphys}         
\begin{document}
\title{Bethe Ansatz and boundary energy of the open spin-$1/2$ XXZ chain}  
%
\authori{Rajan Murgan}      \addressi{Physics Department,P.O.Box 248046,University of Miami Coral Gables,FL 33124 USA}
\authorii{}     \addressii{}
\authoriii{}    \addressiii{}
\authoriv{}     \addressiv{}
\authorv{}      \addressv{}
\authorvi{}     \addressvi{}
%
\headauthor{R.Murgan}            
\headtitle{Bethe Ansatz and boundary energy \ldots}             
\lastevenhead{R.Murgan} 
\refnum{A}
\daterec{13 July 2006}    
\issuenumber{10/11}  \year{2006}
\setcounter{page}{1}
\maketitle

\begin{abstract}
We review recent results on the Bethe Ansatz solutions for the eigenvalues of the transfer matrix of an integrable open XXZ quantum spin chain using functional relations which the transfer matrix obeys at roots of unity.
First, we consider a case where at most two of the boundary parameters 
{{$\alpha_-$,$\alpha_+$,$\beta_-$,$\beta_+$}} are nonzero. A generalization of the Baxter $T-Q$ equation that involves more than one 
independent $Q$ is described. We use this solution to compute the boundary energy of the chain in 
the thermodynamic limit. We conclude the paper with a review of some results for the general 
integrable boundary terms, where all six boundary parameters are arbitrary. 
\end{abstract}

\section{Introduction}
While the open spin-$1/2$ XXZ quantum spin chain (with diagonal boundary terms) has been solved and well understood \cite{Ga, ABBBQ, Sk}, the solution for the  corresponding XXZ chain with general integrable boundary terms, has remained unsolved. The Hamiltonian for this model is given by \cite{dVGR, GZ} 
\bea
{\cal H } & = & \sum_{n=1}^{N-1}{1 \over 2} \left(\sigma_n^x\sigma_{n+1}^x+\sigma_n^y\sigma_{n+1}^y+\cosh \eta \sigma_n^z\sigma_{n+1}^z\right)\nonumber\\
&  & +{1\over 2}\sinh \eta \Big[ 
\coth \alpha_{-} \tanh \beta_{-}\sigma_{1}^{z}
+ \mbox{cosech}\ \alpha_{-} \mbox{sech}\ \beta_{-}\big( 
\cosh \theta_{-}\sigma_{1}^{x}  
+ i\sinh \theta_{-}\sigma_{1}^{y} \big)\nonumber\\
& & - \coth \alpha_{+} \tanh \beta_{+} \sigma_{N}^{z}
+ \mbox{cosech}\ \alpha_{+} \mbox{sech}\ \beta_{+}\big( 
\cosh \theta_{+}\sigma_{N}^{x}
+ i\sinh \theta_{+}\sigma_{N}^{y} \big)
\Big] 
\eea
where $\sigma^x$,$\sigma^y$,$\sigma^z$ are Pauli matrices, $\eta$ is the bulk anisotropy parameter, $\alpha_{\pm}$,$\beta_{\pm}$,$\theta_{\pm}$ are the boundary parameters, and $N$ is the number of spins. However, the case of nondiagonal boundary terms with the boundary parameters satisfying certain constraint 
has been solved recently \cite{Ne2, CLSW}. Hence, it would be desirable to find the solution for the general case, with such constraint removed. \newline \\ The outline of this paper is as follows. In section 2, we review our Bethe Ansatz solutions for special case at roots of unity \cite{MN2} which we utilize to compute the boundary (surface) energy of the XXZ chain \cite{MNS}. Next, in section 3, we present the Bethe Ansatz solution for the general case \cite{MNS2}. This is followed by a brief conclusion of the paper together with some outline of possible future works on the subject in section 4.       
 
\section{Special case}      

Here, the results are presented for odd values of $p$ \footnote{for even $p$ values,we refer readers to \cite{MN1} }, where $p$ is related to $\eta$ through $\eta={i\pi\over p+1}$.

\subsection{Bethe Ansatz}   

We consider the case with the following choice of boundary parameters; $\beta_{\pm} = 0$\,,$\theta_{-}=\theta_{+}$\,,$\alpha_{\pm}$ arbitrary. One crucial step here is to notice that certain functional relation which the transfer matrix, $t(u)$ and its eigenvalues, $\Lambda(u)$ obey at roots of unity \cite{NPB}, can be written as \cite{BR}       
\bea
\det {\mathcal M}(u) = 0 \,,
\label{detzero}
\eea
We give an example of the functional relation below, for $p=3$
\bea
& &\Lambda(u) \Lambda(u+\eta) \Lambda(u+2\eta) \Lambda(u+3\eta) 
- \delta(u) \Lambda(u+2\eta) \Lambda(u+3\eta) 
- \delta(u+\eta) \Lambda(u) \Lambda(u+3\eta) \nonumber \\
& &\quad  -\delta(u+2\eta) \Lambda(u) \Lambda(u+\eta) 
- \delta(u+3\eta) \Lambda(u+\eta) \Lambda(u+2\eta) \nonumber \\
& &\quad +\delta(u) \delta(u+2\eta)
+ \delta(u+\eta) \delta(u+3\eta)
= f(u) \,.
\eea 
$\delta(u)$ and $f(u)$ are known scalar functions in terms of boundary parameters \cite{MN1, MN2}.
The matrix ${\mathcal M}(u)$ is given by \cite{MN2},
\bea
{\mathcal M}(u) = 
\left(
\begin{array}{cccccccc}
    \Lambda(u) & -{\delta(u)\over h^{(1)}(u)} & 0  & \ldots  & 0 &
    -{\delta(u-\eta)\over h^{(2)}(u-\eta)}  \\
    -h^{(1)}(u) & \Lambda(u+\eta) & -h^{(2)}(u+\eta)  & \ldots  & 0 & 0  \\
    \vdots  & \vdots & \vdots & \ddots 
    & \vdots  & \vdots    \\
   -h^{(2)}(u-\eta)  & 0 & 0 & \ldots  & -h^{(1)}(u+(p-1)\eta) &
    \Lambda(u+p\eta) 
\end{array} \right)  \,,
\eea
where $h^{(1)}(u)$ and $h^{(2)}(u)$ are functions which are
$i\pi$-periodic. Comparing (2) to the functional relation for the eigenvalues, one can solve for $h^{(1)}(u)$. Also, $h^{(2)}(u) = h^{(1)}(-u-2\eta)$, as one would conclude from the crossing properties of $\Lambda(u)$ and (8) below.
The matrix above has the following symmetry,  $T {\cal M}(u) T^{-1} = {\cal M}(u+2\eta)$ and $T \equiv S^{2}$. Other ${\mathcal M}(u)$ matrices we found with stronger symmetry yield inconsistent results. Details on this argument can be found in \cite{MN2}.  
Here $S$ is
\bea
S = \left(
\begin{array}{cccccccc}
    0 & 1 & 0  & \ldots  & 0 & 0  \\
    0 & 0 & 1  & \ldots  & 0 & 0  \\
    \vdots  & \vdots & \vdots & \ddots 
    & \vdots  & \vdots    \\
    0 & 0 & 0  & \ldots  & 0 & 1 \\
   1  & 0 & 0 & \ldots  & 0 & 0
\end{array} \right) \,, \qquad S^{p+1} =  I.
\label{Smatrix}
\eea 
Hence, we have
\bea
h^{(1)}(u) &=& {8\sinh^{2N+1}(u+2\eta)\cosh^{2}(u+\eta) \cosh(u+2\eta)\over 
\sinh(2u+3\eta)} \,,
 \eea
The above symmetry for the present ${\cal M}(u)$ suggests a null eigenvector, $v(u)= \left( Q_{1}(u)\,, Q_{2}(u+\eta) \,, \ldots \,, Q_{1}(u-2\eta) 
\,, Q_{2}(u-\eta)\right)$ with the following ansatz for $Q_{a}(u)$  
\bea
Q_{a}(u) = \prod_{j=1}^{M_{a}} 
\sinh (u - u_{j}^{(a)}) \sinh (u + u_{j}^{(a)} + \eta) \,, \qquad a = 
1\,, 2\,, 
\eea
Note that there are two $Q(u)$ functions, a direct consequence of the weaker symmetry mentioned above. Thus, we have the following $T-Q$ relations
\bea
\Lambda(u) &=& 
{\delta(u)\over h^{(1)}(u)} {Q_{2}(u+\eta)\over Q_{1}(u)} 
+ {\delta(u-\eta)\over h^{(2)}(u-\eta)} {Q_{2}(u-\eta)\over 
Q_{1}(u)} \,, \label{TQ1} \nonumber\\
 &=& 
h^{(1)}(u-\eta) {Q_{1}(u-\eta)\over Q_{2}(u)} 
+ h^{(2)}(u) {Q_{1}(u+\eta)\over 
Q_{2}(u)} \,.
\eea
with  $M_{1} = {1\over 2}(N+p+1)$ and  $M_{2} = {1\over 2}(N+p-1) \,.$

We see that (8) is a coupled equation in terms of $Q_{1}(u)$ and $Q_{2}(u)$, hence exhibiting a generalized structure of $T-Q$ relation.
Bethe Ansatz follows directly by demanding analyticity for the $\Lambda(u)$.
\bea
{\delta(u_{j}^{(1)})\ h^{(2)}(u_{j}^{(1)}-\eta)
\over \delta(u_{j}^{(1)}-\eta)\ h^{(1)}(u_{j}^{(1)})} 
&=&-{Q_{2}(u_{j}^{(1)}-\eta)\over Q_{2}(u_{j}^{(1)}+\eta)} \,, \qquad j =
1\,, 2\,, \ldots \,, M_{1} \,, \nonumber \\
{h^{(1)}(u_{j}^{(2)}-\eta)\over h^{(2)}(u_{j}^{(2)})}
&=&-{Q_{1}(u_{j}^{(2)}+\eta)\over Q_{1}(u_{j}^{(2)}-\eta)} \,, \qquad j =
1\,, 2\,, \ldots \,, M_{2} \,.
\eea 

\subsection{Boundary energy}   

The energy for the chain of finite length is given by
\bea
E={1\over 2} \sinh^{2}\eta 
\sum_{a=1}^{2}\sum_{j=1}^{M_{a}}{1\over 
\sinh (\tilde u_{j}^{(a)} - {\eta\over2})
\sinh (\tilde u_{j}^{(a)} + {\eta\over2})} + {1\over 2}(N-1) \cosh 
\eta \,
\eea
where $\tilde u_{j}^{(a)} \equiv u_{j}^{(a)} + {\eta\over2}$, 
We make the string hypothesis that, for suitable values of boundary parameters \footnote{Readers are urged to refer to \cite{MNS} for a detail discussion on this matter}, the ground state roots,$\{ \tilde u_{j}^{(1)} \}$ and $\{ \tilde
u_{j}^{(2)} \}$ have the following form as $N\rightarrow\infty$.
\bea
\left\{ \begin{array}{c@{\quad : \quad} l}
v_{j}^{(1,1)}                      & j = 1\,, 2\,, \ldots \,, {N\over 2} \\
v_{j}^{(1,2)} + {i \pi\over 2} \,, & j = 1\,, 2\,, \ldots \,, {p+1\over 2}
\end{array} \right. \,, \qquad
\left\{ \begin{array}{c@{\quad : \quad} l}
v_{j}^{(2,1)}                      & j = 1\,, 2\,, \ldots \,, {N\over 2} \\
v_{j}^{(2,2)} + {i \pi\over 2} \,, & j = 1\,, 2\,, \ldots \,, {p-1\over 2}
\end{array} \right. \,,
\eea  

\noindent $\{ v_{j}^{(a,b)} \}$ are  all real and positive.
The logarithm of the Bethe equations for both sets of sea roots, $\{v_{j}^{(1,1)} \}$ and $\{
v_{j}^{(2,1)} \}$ gives the ground state root densities, $\rho^{(1)}(\lambda)$ and $\rho^{(2)}(\lambda)$ with $v_{j}^{(a,b)} = \mu \lambda_{j}^{(a,b)}
$.
The energy depends only on the sum of root densities computed from the counting functions. Further, using (10) (where $\sum\ldots\rightarrow $N$\int(\rho^{(1)}(\lambda)+\rho^{(2)}(\lambda))d\lambda\ldots$ in the thermodynamic limit, $N\rightarrow\infty$)
and keeping term of order 1, we have the following
\bea
E_{boundary}^{\pm}&=& - {\sin \mu\over 2\mu} 
\int_{-\infty}^{\infty} d\omega\ 
{1\over 2\cosh (\omega/ 2)}
\Big\{ 
{\cosh((\nu-2)\omega/4) \over 2\cosh(\nu \omega/4)} 
-{1\over 2} \nonumber\\
&+& {\sinh(\omega/2) \cosh ((\nu-2|a_{\pm}|)\omega/2) \over 
\sinh(\nu \omega/2)} \Big\} -{1\over 4}\cos \mu \,.
\eea
where $\alpha_{\pm}=i\mu a_{\pm}$ and $\mu=-i\eta$. + and - refer to right and left boundary respectively.

\section{General case ($p>1$)}      

Finally, we present the solution for the case of general nondiagonal boundary terms. We first present the matrix, ${\cal M}(u)$, for this case
\bea
{\cal M}(u) = \left(
\begin{array}{ccccccccc}
    \Lambda(u) & -m_{1}(u) & 0  & \ldots  & 0 & 0 & -n_{p+1}(u)  \\
    -n_{1}(u) & \Lambda(u+\eta) & -m_{2}(u)  & \ldots & 0 & 0 & 0  \\
    \vdots  & \vdots & \vdots & \ddots 
    & \vdots  & \vdots  & \vdots   \\
      0 & 0 & 0 & \ldots  & -n_{p-1}(u) &
       \Lambda(u+(p-1) \eta) & -m_{p}(u) \\
      -m_{p+1}(u)  & 0 & 0 & \ldots & 0 & -n_{p}(u) &
    \Lambda(u+p \eta)
\end{array} \right)  
\eea
where the matrix elements,$\{ m_{j}(u) \,,  n_{j}(u) \}$ are given by \cite{MNS2} 
\bea
m_{j}(u) &=& h(-u-j \eta) \,, \qquad  n_{j}(u) = h(u+j \eta) \,, \qquad
j=1\,, 2\,, \ldots \,, p \,, \nonumber \\
m_{p+1}(u) &=& {z^{-}(u) \over 
\prod_{k=1}^{p}h(-u-k\eta)} \,, \qquad n_{p+1}(u) = {z^{+}(u) \over 
\prod_{k=1}^{p}h(u+k\eta)}\,, 
\eea
where
\bea
h(u) = -4\sinh^{2N}(u+\eta){\sinh(2u+2\eta)\over \sinh(2u+\eta)}\times\nonumber\\
       \sinh(u+\alpha_{-}) \cosh(u+\beta_{-})\sinh(u+\alpha_{+}) \cosh(u+\beta_{+}) 
\eea
and
\bea
z^{\pm}(u)={1\over 2}\left(f(u) \pm g(u)\ Y(u) \right) 
\eea 
Explicit expressions for $g(u)$ and $Y(u)$ (a non-analytic function) and their properties are given in \cite{MNS2}. 
The null eigenvector is $v(u) = (v_{1}(u) \,, v_{2}(u)\,, \ldots  \,, v_{p+1}(u))$\footnote{$v_{j+p+1} = v_{j}$ }.
Periodicity of ${\cal M}(u)$ makes it reasonable to assume the same $i\pi$ periodicity for $v(u)$. Utilizing ${\cal M}(u)v(u)=0$ together with the following ansatz for $v_{j}(u)$ \footnote{with following crossing property, $v_{j}(-u) = v_{p+2-j}(u)$ \,, \qquad j = 1\,, 2\,, \ldots \,, p+1 \,.},  
\bea
v_{j}(u) = a_{j}(u) + b_{j}(u)\ Y(u) \,, \qquad j = 1\,, 2 \,, \ldots 
\,, \lfloor{p\over 2}\rfloor +1 \,, 
\eea 
where
\bea
a_{j}(u) &=& A_{j} \prod_{k=1}^{2M_{a}} \sinh(u-u_{k}^{(a_{j})}) \,, \qquad 
b_{j}(u) = B_{j}\prod_{k=1}^{2M_{b}} \sinh(u-u_{k}^{(b_{j})})\,,
\qquad j \ne {p\over 2}+1 \,, \nonumber \\
a_{{p\over 2}+1}(u) &=& A_{{p\over 2}+1} 
\prod_{k=1}^{M_{a}} \sinh(u-u_{k}^{(a_{{p\over 2}+1})})
\sinh(u+u_{k}^{(a_{{p\over 2}+1})}) \,, \nonumber \\
b_{{p\over 2}+1}(u) &=& B_{{p\over 2}+1}
\prod_{k=1}^{M_{b}} \sinh(u-u_{k}^{(b_{{p\over 2}+1})}) 
\sinh(u+u_{k}^{(b_{{p\over 2}+1})}) \,, 
\eea
 and equating analytic and non-analytic terms separately, one would derive a set of generalized $T-Q$ equations \cite{MNS2}. 
 Here  $M_{a} = \lfloor {N-1\over 2} \rfloor + 2p+1$ and  $M_{b} = \lfloor {N-1\over 2} \rfloor + p$.

Using similar arguments as in section 2.1, and invoking analyticity of $\Lambda(u)$, we get the following Bethe-Ansatz like equations for the zeros $\{ u_{l}^{(a_{j})} \}$ and $\{ u_{l}^{(b_{j})} \}$ of
the functions $\{ a_{j}(u)\}$ and $\{ b_{j}(u)\}$ respectively,
\bea
h(-u_{l}^{(a_{1})}-\eta)
&=&-{f(u_{l}^{(a_{1})})\ a_{1}(-u_{l}^{(a_{1})}) +
g(u_{l}^{(a_{1})})\ Y(u_{l}^{(a_{1})})^{2}\ 
b_{1}(-u_{l}^{(a_{1})})\over
2a_{2}(u_{l}^{(a_{1})})\
\prod_{k=1}^{p}h(u_{l}^{(a_{1})}+k\eta)} \,, \nonumber \\
{h(-u_{l}^{(a_{j})}-j\eta)\over 
h(u_{l}^{(a_{j})}+(j-1)\eta)}&=&-{a_{j-1}(u_{l}^{(a_{j})})\over
a_{j+1}(u_{l}^{(a_{j})})}  \,, \qquad 
j = 2 \,, \ldots \,, \lfloor{p\over 2}\rfloor+1 \,,
\eea
and
\bea
h(-u_{l}^{(b_{1})}-\eta)
&=&-{f(u_{l}^{(b_{1})})\ b_{1}(-u_{l}^{(b_{1})}) +
g(u_{l}^{(b_{1})})\  
a_{1}(-u_{l}^{(b_{1})})\over
2b_{2}(u_{l}^{(b_{1})})\ 
\prod_{k=1}^{p}h(u_{l}^{(b_{1})}+k\eta)} \,, \nonumber \\
{h(-u_{l}^{(b_{j})}-j\eta)\over 
h(u_{l}^{(b_{j})}+(j-1)\eta)}&=&-{b_{j-1}(u_{l}^{(b_{j})})\over
b_{j+1}(u_{l}^{(b_{j})})}  \,, \qquad 
j = 2 \,, \ldots \,, \lfloor{p\over 2}\rfloor+1 \,.
\eea
Normalization contants $\{ A_{j}, B_{j} \}$\,,$j=1 \,, \ldots \,,\lfloor{p\over 2}\rfloor+1 \,,$ can be determined by noting the poles at $u = -{\eta\over 2}$ and $u = -\alpha_{-}-\eta$, and from the analyticity of $\Lambda(u)$. This yields few extra Bethe-Ansatz like equations that can be solved for these normalization constants. 

\section{Conclusion}      
\indent
We have presented Bethe Ansatz solutions for both special and general cases. The solutions presented here have been verified for completeness numerically. However, these solutions do not hold for generic values of $\eta$, but only for special values, ${i \pi\over p+1}$. Further, we have demonstrated the use of string hypothesis to compute the ground state boundary energy for a special case.
The Bethe Ansatz equations appear in a generalized form due to the appearance of multiple $Q(u)$ (or $a(u)$ and $b(u)$). 
Few questions needed to be answered here. Firstly, having found the general solutions, can they lead to some other interesting results, e.g. finite size effects? Secondly, do solutions exist for generic values of $\eta$?  These are certainly questions of utmost importance that we hope to pursue and address in future.    

\section*{Acknowledgments}

I would like to thank R. I. Nepomechie for his invaluable help and advice during the course of completing this work. I also thank the Department of Physics and Graduate School (Arts and Sciences), University of Miami for their financial support, and C. Shi for his useful suggestions and comments.
This work was supported in part by the NSF under Grant PHY-0244261.

\end {document}